# Tailoring excitonic states of van der Waals bilayers through stacking configuration, band alignment and valley-spin

Wei-Ting Hsu,[1] Bo-Han Lin,[2] Li-Syuan Lu,[2] Ming-Hao Lee,[3] Ming-Wen Chu,[3] Lain-Jong Li,[4] Wang Yao,[5,*] Wen-Hao Chang,[2,*] and Chih-Kang Shih[1,*]

[1] Department of Physics, The University of Texas at Austin, Austin, Texas 78712, USA

[2] Department of Electrophysics, National Chiao Tung University, Hsinchu 30010, Taiwan

[3] Center for Condensed Matter Sciences, National Taiwan University, Taipei 10617, Taiwan

[4] Corporate Research and Chief Technology Office, Taiwan Semiconductor Manufacturing Company (TSMC), Hsinchu 30075, Taiwan

[5] Department of Physics and Center of Theoretical and Computational Physics, The University of Hong Kong, Hong Kong, China

*Correspondence and requests for materials should be addressed to:

shih@physics.utexas.edu (C.K.S.)

whchang@mail.nctu.edu.tw (W.H.C.)

wangyao@hku.hk (W.Y.)

**Excitons in monolayer semiconductors have large optical transition dipole for strong coupling with light field. Interlayer excitons in heterobilayers, with layer separation of electron and hole components, feature large electric dipole that enables strong coupling with electric field and exciton-exciton interaction, at the cost that the optical dipole is substantially quenched (by several orders of magnitude). In this letter, we demonstrate the ability to create a new class of excitons in transition metal dichalcogenide (TMD) hetero- and homo-bilayers that combines the advantages of monolayer- and interlayer-excitons, i.e. featuring both large optical dipole and large electric dipole. These excitons consist of an**

**electron that is well confined in an individual layer, and a hole that is well extended in both layers, realized here through the carrier-species specific layer-hybridization controlled through the interplay of rotational, translational, band offset, and valley-spin degrees of freedom. We observe different species of such layer-hybridized valley excitons in different heterobilayer and homobilayer systems, which can be utilized for realizing strongly interacting excitonic/polaritonic gases, as well as optical quantum coherent controls of bidirectional interlayer carrier transfer either with upper conversion or down conversion in energy.**

Van der Waals (vdW) bilayers composed of stacking two atomically-thin 2D layers have rapidly evolved into a variety of designer quantum materials with many fascinating properties that are not possessed by the constituent monolayers [1-12]. The key control knob for designing new quantum materials is the stacking dependent interlayer coupling, which determines the electronic structure of the vdW bilayer as a whole. One strategy is to harness the *"moiré pattern"* — a periodic variation of local stacking configuration (due to twist angle and/or lattice mismatch) — to form a 2D electronic superlattice. Such a *"moiré designer"* has led to the observation of the Mott insulating phase, and more intriguingly, the accompanying unconventional superconductivity in twisted bilayer graphene [13,14]. In semiconducting transition metal dichalcogenide (TMD) hetero-bilayers, this strategy has also been utilized to create interlayer *moiré excitons* which possess attractive properties such as tunable quantum emitters and spin-orbit coupled superlattice [12,15-17]. While the formation of the moiré exciton has been stated, an in situ *structural* confirmation is still needed to directly connect the observed excitonic features to the formation of excitonic moiré superlattice in the experiments [15-17].

In this letter we report structural and spectroscopic evidences that coherently show how stacking configurations, band alignment, and valley-specific spin configurations work in concert to define the excitonic structures of the vdW bilayers. More specifically, utilizing the above degrees of freedom, we tailor different species of layer-hybridized excitonic states in TMD hetero- and homo-bilayers that feature both large optical dipole (comparable to that of monolayer exciton) and large electric dipole (comparable to that of interlayer exciton), which are highly desired for realizing strongly interacting excitonic/polaritonic gases and can also be exploited for spin-valley selective interlayer quantum controls. In addition, we determine quantitatively important coupling parameters for future design of electronic structures of vdW bilayers.

The layer-hybridization of electronic states in bilayers can be described by a two-level Hamiltonian: $\begin{bmatrix} \varepsilon_u & -t \\ -t & \varepsilon_l \end{bmatrix}$ (**Fig. 1a**), where the basis states $|\psi_u\rangle$ and $|\psi_l\rangle$ are the states from upper and lower layers before coupling. The off-diagonal element $t$ is the interlayer hopping integral that conserves the spin. For the electronic states at the $\pm K$ valleys, the hopping integral $t$ is strongly dependent on the stacking configuration [11]. Noticeably at high symmetry stacking, 3-fold rotational symmetry dictates $t$ to be zero/finite at certain band-edges, where interlayer hybridization is forbidden/allowed (c.f. **Fig. 1b-f**). On the other hand, the offset $2\delta = \varepsilon_l - \varepsilon_u$ between the band edges from the two layers is controlled by valley-spin splitting together with the valley pairing that is opposite in the R- and H-type stackings. The two layer-hybridized eigen vectors are $|\psi_{h+}\rangle = \alpha|\psi_u\rangle + \beta|\psi_l\rangle$ and $|\psi_{h-}\rangle = \alpha'|\psi_u\rangle + \beta'|\psi_l\rangle$, where $\beta/\alpha = t/(\delta - \Delta)$, $\beta'/\alpha' = t/(\delta + \Delta)$ and $\Delta = \sqrt{\delta^2 + t^2}$. The energy separation between the doublet is $2\Delta$, which, together with the spectral ratio $|\beta/\alpha|^2$, can provide signature of the hybridization. The degree of layer-hybridization is further defined as $P_H = |\beta/\alpha|$.

Our vdW bilayer samples include WSe$_2$/MoSe$_2$ and WS$_2$/MoS$_2$ hetero-bilayers with negligible lattice mismatch, and MoS$_2$ homo-bilayers grown directly by chemical vapor deposition (CVD) [10]. The fundamental characterizations can be found in **Supplementary Fig. S1-3**. Despite a slight lattice mismatch of $\sim 0.2 - 0.4\%$ between WX$_2$ and MoX$_2$ (X = S, Se), the top layer of the CVD-grown WX$_2$/MoX$_2$ hetero-bilayers tends to adopt the same lattice constant as the bottom layer (**Supplementary Fig. S4**) to form commensurate R (twist angle $\theta = 0°$) and H stacking ($\theta = 60°$), as schematically shown in **Fig. 1b** and **1c**. The interlayer atomic registry has been examined by transmission electron microscopy (TEM). Only the $H_X^M$ stacking configuration was observed in H stacking samples. TEM inspections on different locations of the same sample also confirm that the stacking has no interlayer translation [10]. In R stacking samples, only the $R_X^M$ or $R_M^X$ stacking configurations were observed (**Fig. 2a-b**). According to **Fig. 1f**, the hopping integral reaches the maximum in $H_X^M$ stacking but vanishes in $R_X^M$ and $R_M^X$ stacking configurations.

The stacking dependent interlayer coupling, the band alignment and the valley-spin work collaboratively to determine the hybridization of electronic states at K point, leading to rich excitonic structures. Shown in **Fig. 2c** are the differential reflectance (DR) spectra acquired at 4K for a $H_X^M$ stacked and a $R_M^X$ stacked WSe$_2$/MoSe$_2$ hetero-bilayer. The most significant difference emerges from the B exciton of WSe$_2$ (denoted as $X_B^W$), which shows a single peak in R stacking, but becomes a doublet (referred to as $X_{h+}^W$ and $X_{h-}^W$) in H stacking with a splitting of 92 meV and a spectral weight ratio of roughly 1:1. In addition, the oscillator strength of the $X_A^{Mo}$ ($X_A^W$) excitons is reduced (enhanced) in H stacking. This can be explained by the splitting of the $X_A^{Mo}$ exciton peak into a doublet ($X_{h+}^{Mo}$ and $X_{h-}^{Mo}$), where the split off spectral feature merges with the $X_A^W$ excitons. Curve fitting yields a splitting of 87 meV and a spectral weight ratio of nearly 1:1.

These spectral features can be further enhanced by taking the second derivative of these spectra, as displayed in **Fig. 2c**. The measurements have been carried out on several samples with well-defined R or H stacking. In all cases, the $H_X^M$ stacked hetero-bilayers exhibit spectral splitting in $X_B^W$, albeit the splitting varies from 70 to 105 meV (**Supplementary Fig. S5**). In **Fig. 2d** we show an example of spectral fitting for an H-type hetero-bilayer.

The $WSe_2/MoSe_2$ hetero-bilayer is known to exhibit a type-II band alignment, where the valence band maximum (VBM) of $WSe_2$ is lying above that of $MoSe_2$, as schematically shown in **Fig. 2e**. We label the two valence bands split by spin-orbit coupling (SOC) as v1 and v2 for $WSe_2$, and v1' and v2' for $MoSe_2$. Due to the larger SOC splitting in $WSe_2$ and the type-II alignment, the v2 band aligns closer to the v1' and v2' bands of $MoSe_2$ [18]. Spin conservation of interlayer hopping, on the other hand, ensures that the v2 band of $WSe_2$ only coupled to the v1' band of $MoSe_2$ in H stacking (**Fig. 2e**). Such a coupling hybridizes v2 and v1' bands into the doublet h+ and h-, as depicted in **Fig. 2e**. Conduction band hybridization, however, is not allowed in this high symmetry stacking [11]. Interband optical transition can excite either of the hybridized valance states (h+ and h-) to the conduction band localized in the $WSe_2$ ($MoSe_2$) layer, leading to the splitting of the $X_B^W$ ($X_A^{Mo}$) exciton state into a doublet $X_{h+}^W$ and $X_{h-}^W$ ($X_{h+}^{Mo}$ and $X_{h-}^{Mo}$) as shown in **Fig. 2e**. These four exciton species, all consisting of a layer hybridized hole and an electron confined in an individual layer (**Fig. 2f**), feature a large optical dipole that is one-half of the monolayer exciton one as well as a large electric dipole in the out-of-plane direction. They therefore combine the advantage of both intralayer excitons (for strong light coupling) and interlayer excitons (for electric tunability of resonances and strong dipole-dipole interaction).

According to the coupled two-level model, the band offset $2\delta$ and the hopping integral $t$ can be deduced from the measured energy splitting ($2\sqrt{\delta^2 + t^2}$) and the spectral weight ratio $(|\beta/\alpha|^2)$. Measurements conducted on several samples show that the spectral weight ratio is in the range of 1.1 to 1.5 and the splitting is ranging from 70 to 105 meV, which determines a hopping integral of $t_{vv} = 43 \pm 9$ meV and a band offset of $2\delta = 12 \pm 2$ meV between the v2 and v1' bands. We thus obtained a very high degree of layer-hybridization of $P_H \approx 87\%$ for holes. We have also performed temperature dependent measurements from 4 to 300K. We found that the splitting and spectral ratio are temperature independent, indicating that the band offset and interlayer hopping integral remain invariant in this temperature range (**Supplementary Fig. S6**).

More importantly, the layer-hybridized excitonic states further enable the possibility of interlayer quantum control. As depicted in **Fig. 2e** and **2g**, the excitonic transitions form a three-level $\Lambda$-system that would allow the interlayer quantum control of electrons via the layer-hybridized holes. The schematic showing in **Fig. 2g** is an example of two layer-confined electron states in MoSe$_2$ ($|\psi_e^{Mo}\rangle$) and WSe$_2$ ($|\psi_e^W\rangle$) layer, which are connected by an interlayer negative trion (IX$^-$). Using X$_{h+}^W$ and X$_{h+}^{Mo}$ optical transitions, the coherent transfer of an electron state from lower MoSe$_2$ to upper WSe$_2$ layer, or vice versa, is therefore possible by the intermediate trion state. Note that this control can be further made into a spin-valley selection by the circular polarization of the light field.

Similar interlayer hybridization has also been observed in WS$_2$/MoS$_2$ hetero-bilayers at room temperature. For this sample set, we also performed measurements on samples with different twist angles. We have identified several commensurate hetero-bilayers with the $R_M^X$, $R_X^M$ and $H_X^M$ stacking configurations based on TEM analyses (**Fig. 3a-c**) and some twisted bilayers using second harmonic generation (SHG) microscopy. **Figure 3d** shows the second derivative spectra for hetero-bilayers with different twist angles. The doublet of the X$_B^W$ exciton is observed only for $\theta = 60°$ (H stacking). The splitting is smaller ($74 \pm 6$ meV for the spectrum shown in **Fig. 3e**) than that of WSe$_2$/MoSe$_2$ hetero-bilayers, indicative of a weaker interlayer hopping in WS$_2$/MoS$_2$ hetero-bilayers. From the measured splitting and fitted spectral weight, we determine the interlayer hopping to be $t_{vv} = 36 \pm 4$ meV, band offset $2\delta = 17 \pm 5$ meV and $P_H \approx 79\%$ in WS$_2$/MoS$_2$ hetero-bilayers. For the $R_M^X$ and $R_X^M$ hetero-bilayers (**Fig. 1f** and **3a-b**), layer-hybridized valley excitons are absent since the interlayer hopping is dictated to be 0 for both the conduction and valance band edges. On the other hand, the absence of interlayer hybridization in twisted bilayers can be understood from two factors. First, the K valleys of two layers in momentum space are misaligned by the twist angle $\theta$ (**Fig. 3d** inset), where the interlayer hopping is inhibited by the large momentum mismatch. Second, since the interlayer hopping decreases exponentially with the interlayer spacing $d$, the enlarged interlayer spacing in twisted hetero-bilayers thus further suppresses the interlayer hopping [19-22].

The $H_X^M$ hetero-bilayers demonstrated above exemplify the interlayer hybridization in the regime with $t > \delta$. The $R_M^X$ and $R_X^M$ hetero-bilayers, on the other hand, represent the example of vanished interlayer hopping $t = 0$ [11]. In principle, one can tune the band offset $2\delta$ by a vertical electric field to tune the degree of hybridization [23]. However, fabricating the device to achieve such tunability is not straightforward, especially when the hetero-bilayer area is small. Here we show that H stacked MoS$_2$ homobilayers can be a model system to investigate the interlayer hybridization of valance bands in the regime with $\delta > t$. The MoS$_2$ bilayers investigated here include commensurate $R_M^X$ and $H_X^M$ stacking and twisted bilayers, as characterized by TEM and SHG measurements (**Supplementary Fig. S7**). **Figure 4a** shows the second derivative spectra for bilayer MoS$_2$ as a function of $\theta$, showing A-exciton ($X_A^{Mo}$) and B-exciton ($X_B^{Mo}$) transitions in all samples. Due to the small spin splitting in the conduction band of MoS$_2$, the energy splitting between $X_A^{Mo}$ and $X_B^{Mo}$ thus represents a good measure of the valence-band spin splitting at the K valley. As depicted in **Fig. 4b**, the A-B exciton splitting for $R_M^X$ and twisted bilayers are around 144 meV, which is very close to the value of 148 meV for monolayer MoS$_2$ and insensitive to the twist angle. The slightly reduced energy splitting in bilayers could be caused by the enhanced dielectric screening. On the contrary, the energy splitting of $H_X^M$ bilayer is increased to 164 meV. The same A-B exciton separation between twisted and $R_M^X$ bilayers is a confirmation on the absence of interlayer hybridization in $R_M^X$ bilayer, as discussed above. On the other hand, the enlarged energy splitting indicates the presence of partial interlayer hybridization in $H_X^M$ homobilayer. **Figure 4c** depicts the optical transitions at the K valley of H-type bilayer. Here we denote the valence band spin splitting of monolayer MoS$_2$ as $2\lambda$. The presence (absence) of interlayer hopping $t$ thus modifies the spin splitting to $2\sqrt{\lambda^2 + t^2}$ ($2\lambda$) for $H_X^M$ ($R_M^X$) bilayer. Comparing the A-B exciton separation in $H_X^M$ bilayers and in twisted bilayers without interlayer hopping, we determine the interlayer hopping integral in MoS$_2$ bilayer to be $t_{vv} = 39$ meV, which is close to theoretical calculation [6] and similar to those for WX$_2$/MX$_2$ hetero-bilayers. From the measured $\lambda$ and $t$, the degree of layer-hybridization of hole is obtained to be $P_H \approx$ 26%, which is much smaller than that of hetero-bilayers and forms a partially layer-hybridized hole.

We note the fact that, in bilayer MoS$_2$, the interlayer hybridization enables total eight transitions of partially layer-hybridized valley excitons as shown in **Fig. 4d-e**, with both spin-up and spin-down holes becoming relevant in each valley. Interestingly, four of them $X_h^o$ have larger optical dipoles (~97% compared to that of monolayer exciton) while four of them $X_h^e$ have larger electric dipoles (~97% compared to that of interlayer exciton). The $X_h^o$ and $X_h^e$ exciton transition energies differ by the conduction band spin splitting, not resolvable here due to the small splitting in MoS$_2$ (~3 meV), while the much larger splitting in other TMDs shall allow separate access of these exciton states. Most importantly, two Λ-shape level schemes are enabled in the spin-up and spin-down subspaces respectively by these layer-hybridized excitons, allowing the interlayer quantum control of both spin species in each valley.

Our work has demonstrated the first observation of the interlayer hybridization of K valleys in TMD hetero- and homo-bilayers, which is consistent with the symmetry-dictated registry dependence. The interlayer hopping integral of valence band is determined to be $t_{vv} \approx 36 - 43$ meV, depending on the material combinations. As most focusing researches have been made on the band-edge moiré excitons in TMD hetero-bilayers, the obtained interlayer hopping strength at high-symmetry points provides a measure for the upper limit of confinement potential in a TMD-based moiré superlattice. By using the interlayer hybridization, creating a moiré potential up to ~100 meV becomes feasible in TMD hetero- and homo-bilayers. Moreover, our work points out a more significant moiré modulation effect in electronic structures, where the layer-distribution of out-of-plane wavefunction can strongly depend on in-plane locations in the moiré. It can become a brand new control knob to engineer excitons in the bilayer, paving the way toward the next-generation artificial platform for exploring exciton physics and engineering moiré quantum dot array. In addition, the layer-hybridized valley excitons can be utilized for realizing strongly interacting excitonic/polaritonic gases, as well as spin-valley selective optical quantum coherent controls of bidirectional interlayer carrier transfer either with upper conversion or down conversion in energy.

**Methods**

**Growth of bilayer TMD heterostructures and homostructures.** Single-crystal $WSe_2/MoSe_2$ hetero-bilayers, $WS_2/MoS_2$ hetero-bilayers and $MoS_2$ homo-bilayers were grown on sapphire substrates in a horizontal hot-wall CVD chamber using the one-pot synthesis process [24,25]. The high-purity $WO_3$ (99.995%, Aldrich), $MoO_2$ (99%, Aldrich), Se (99.5%, Alfa) and S (99%, Aldrich) powders were used as the source precursors. The sapphire substrate and metal-oxide powder were placed at the central heating zone, while the chalcogen powder was heated by a heating belt at the upstream end. For $WSe_2/MoSe_2$ hetero-bilayers, the heterostructures were grown at 880℃ in $Ar/H_2$ flowing gas with the flow rates of 60/6 sccm at low pressure (5-40 Torr) [10]. For $WS_2/MoS_2$ hetero-bilayers, the growth temperature was set to 920℃ in $Ar/H_2$ flowing gas with the flow rates of 60/6 sccm at low pressure condition (5 Torr). The $MoS_2$ homo-bilayers were grown at 650℃ in Ar flowing gas at ambient pressure.

**TEM characterizations.** Annular dark field scanning TEM imaging was performed in a spherical aberration corrected transmission electron microscope (JEOL-2100F). The CVD-grown samples were transferred onto the TEM grids using the conventional wet-transfer process. The TMD/sapphire samples were capped with a layer of poly(methylmethacrylate) (PMMA) (950K A4) by spin-coating, followed by baking at 100℃ for 60 min. The PMMA-capped sample was then immersed into a BOE solution at 80℃ for 20 min. After diluting etchant and residues in deionized water, the PMMA film was exfoliated from the sapphire substrate and transferred onto a Cu grid with carbon nets (Ted Pella). Then the top PMMA film was removed by acetone, and the sample was cleaned by isopropyl alcohol and deionized water.

**Optical measurements.** Room-temperature optical characterizations, such as photoluminescence (PL), Raman and SHG spectroscopies were performed using a back-scattering optical microscope. The light sources were focused on the sample by a 100× objective lens (N.A.= 0.9), and the signal was sent to a 0.75 m monochromator, then detected by a nitrogen-cooled CCD camera. For PL and Raman measurements, a 532 nm solid-state laser was used as the excitation source. For SHG measurements, the fundamental laser field was provided by a mode-locked Ti:Sapphire laser at 880 nm. The polarization of fundamental laser (SHG signal) was selected (analyzed) by the individual linear polarizers and half-wave plates [26].

For low-temperature DR measurements, the sample was cooled down to $T$ = 4K by a cryogen-free low-vibration cryostat equipped with a 3-axis piezo-positioner and a 50× objective lens (N.A.= 0.82). A fiber-coupled tungsten-halogen lamp was used as the light source. In order to improve the spatial resolution, the confocal optics was set up in front of the monochromator, resulting in a final spatial resolution of ~0.5 μm. The integration time per spectra is around 4 s, where the signal-to-noise ratio is further improved by averaging > 100 spectra. The second derivative spectra are numerically smoothed using the Savitzky-Golay method, resulting in an overall energy resolution of ~5 meV.

# End Notes

## Acknowledgments

This research was supported with grants from the Welch Foundation (F-1672), the US National Science Foundation (DMR-1808751), the NSF MRSEC program (DMR-1720595) and the US Airforce (FA2386-18-1-4097). W.T.H. acknowledges the support from the Ministry of Science and Technology of Taiwan (MOST-107-2917-I-564-010). W.H.C acknowledges the support from the Ministry of Science and Technology of Taiwan (MOST-105-2119-M-009-014-MY3, MOST-107-2112-M-009-024-MY3). W.Y. acknowledges the support from Research Grants Council of HKSAR (HKU17312916).

## Author contributions

C.K.S. and W.T.H. conceived the idea and designed the experiment. W.T.H and B.H.L. performed the spectroscopy measurements and analyses. The CVD samples were grown and characterized by L.S.L., under the supervision of L.J.L. and W.H.C. M.H.L., W.T.H. and L.S.L. performed the TEM measurements, assisted by M.W.C. W.T.H., W.H.C., W.Y. and C.K.S. developed the model to interpret the spectroscopic data. C.K.S. and W.T.H. wrote the paper with key inputs from W.H.C. and W.Y. All authors discussed the results and commented on the manuscript.

## Additional information

Supplementary information is available in the online version of the paper. Correspondence and requests for materials should be addressed to C.K.S. (shih@physics.utexas.edu), W.H.C. (whchang@mail.nctu.edu.tw) and W.Y. (wangyao@hku.hk).

## Competing financial interests

The authors declare no competing financial interests.

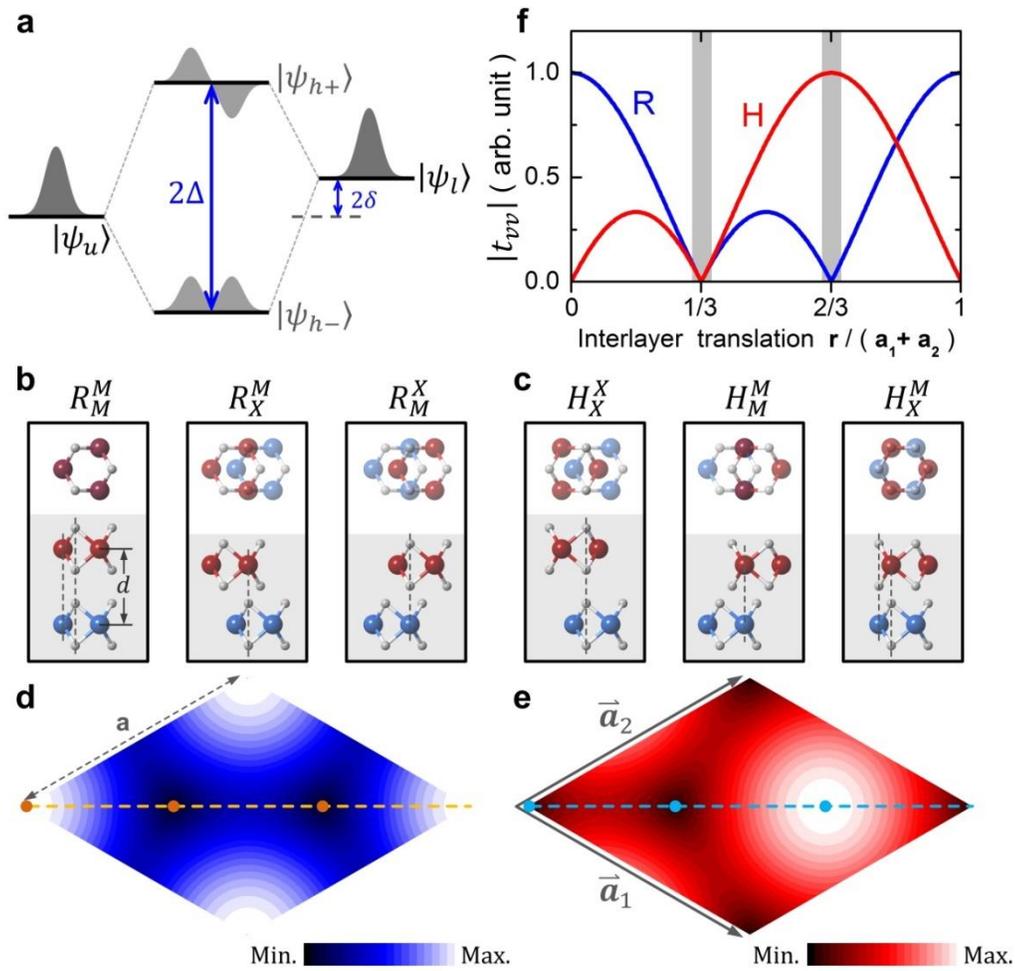

**Figure 1| Interlayer hybridization of valence bands at K valleys for commensurate TMD hetero-bilayer. (a)** A schematic showing the hybridization of electronic states from upper ($|\psi_u\rangle$) and lower ($|\psi_l\rangle$) layers. **(b,c)** High-symmetry R-type (b) and H-type (c) stacking configurations, where the top (side) view is shown in the upper (lower) panel. $R_\nu^\mu$ ($H_\nu^\mu$) denotes an R-type (H-type) stacking with $\mu$-sites of upper layer vertically aligned with $\nu$-sites of lower layer, where $\mu, \nu = M$ or $X$. **(d,e)** The interlayer hopping integral $|t_{\nu\nu}|$ as a function of interlayer translation **r** for R-type (d) and H-type (e) hetero-bilayers, where the dots correspond to the high symmetry stacking configurations shown in (b) and (c), respectively. **(f)** $|t_{\nu\nu}|$ of R-type (blue) and H-type (red) hetero-bilayers along the dashed diagonal line are shown for comparison.

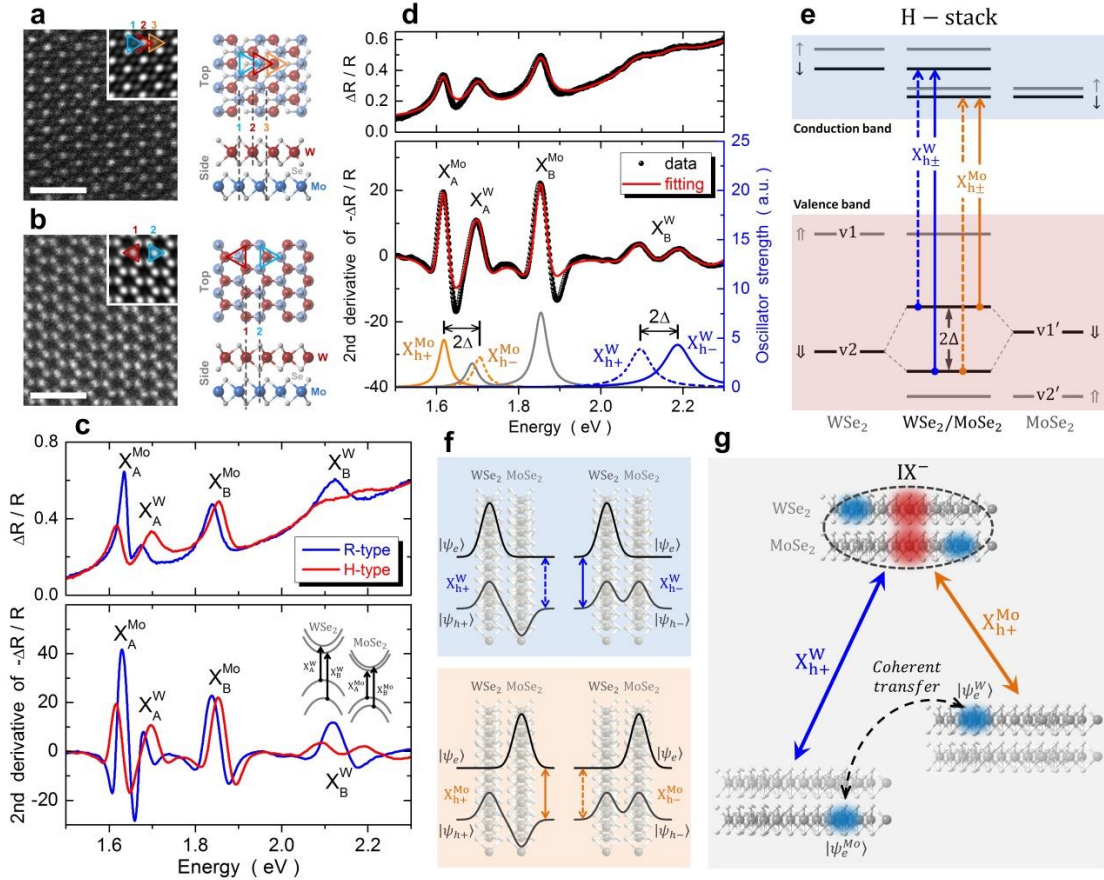

**Figure 2| Layer-hybridized valley excitons in WSe₂/MoSe₂ hetero-bilayer. (a,b)** TEM and Bragg-filtered (inset) images of $R_M^X$ (a) and $H_X^M$ (b) hetero-bilayers. The ideal atomic registry is illustrated for comparison, where the top (side) view is shown in the upper (lower) panel. The scale bar is 1 nm. **(c)** The DR (upper panel) and second derivative (lower panel) spectra of a $R_M^X$ (blue curves) and an $H_X^M$ (red curves) hetero-bilayers, showing a clear splitting of $X_B^W$ for the $H_X^M$ hetero-bilayer. **(d)** The DR (upper panel) and second derivative (lower panel) spectra of an $H_X^M$ hetero-bilayer shown with the spectra fitting (red curve). **(e)** Schematic showing the optical transitions of the $H_X^M$ hetero-bilayer at K valley. Note that both $X_B^W$ (blue) and $X_A^{Mo}$ (orange) transitions are split into two hybridized transitions ($X_{h\pm}^W$ and $X_{h\pm}^{Mo}$), corresponding to the fitting curves in (d). The transitions form Λ-shape level schemes that allow the interlayer quantum control of electrons. **(f)** Schematics showing the wavefunctions in the out-of-plane direction for the four species of layer-hybridized excitons. **(g)** A schematic showing the spin-valley selective

interlayer quantum control of electron states by $X_{h+}^{W}$ and $X_{h+}^{Mo}$ transitions intermediated via an interlayer negative trion $IX^{-}$.

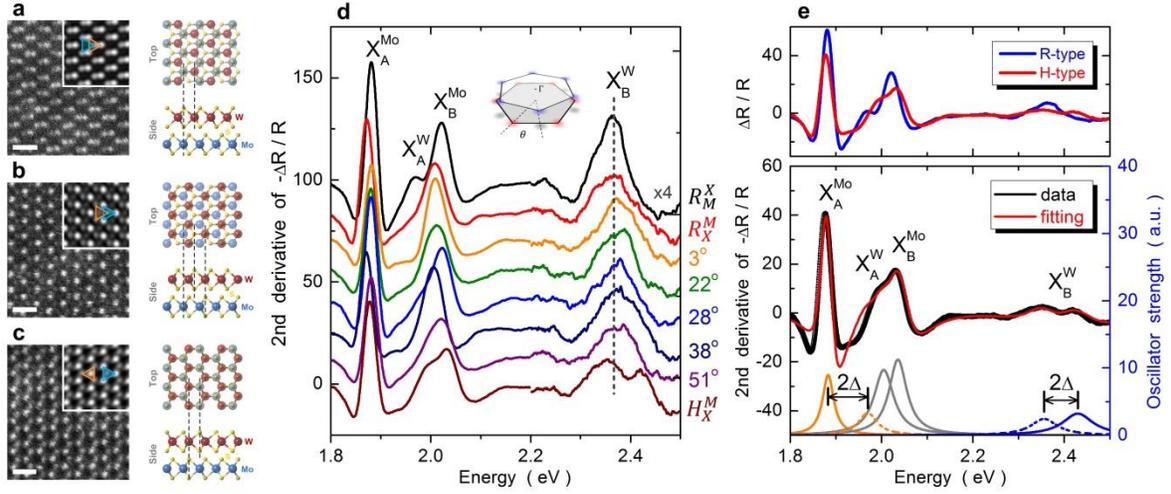

**Figure 3| Layer-hybridized valley excitons in WS₂/MoS₂ hetero-bilayer. (a-c)** TEM and Bragg-filtered (inset) images of $R_M^X$ (a), $R_X^M$ (b) and $H_X^M$ (c) hetero-bilayers. The ideal atomic registry is illustrated for comparison, where the top (side) view is shown in the upper (lower) panel. The scale bar is 0.5 nm. **(d)** The second derivative spectra as a function of twist angle $\theta$. Only the $H_X^M$ hetero-bilayer exhibits interlayer hybridization. **(e)** A comparison of $R_M^X$ and $H_X^M$ hetero-bilayers (upper panel), showing a clear splitting of $X_B^W$ for the $H_X^M$ hetero-bilayer. The second derivative spectrum (lower panel) of the $H_X^M$ hetero-bilayer with spectral fitting (red curve).

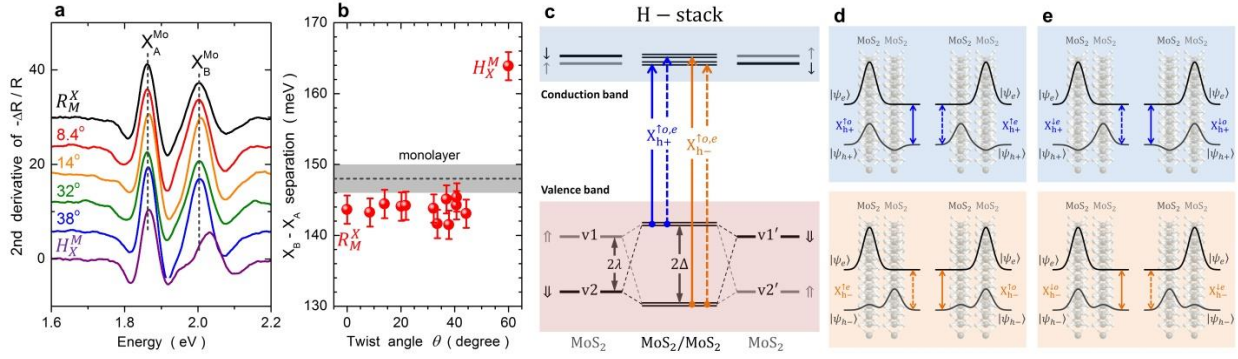

**Figure 4| Partially layer-hybridized valley excitons in bilayer MoS₂.** **(a)** The second derivative spectra of bilayer MoS₂ as a function of twist angle $\theta$. **(b)** The energy separation between $X_A^{Mo}$ and $X_B^{Mo}$ as a function of $\theta$, where the $H_X^M$ bilayer is a singular point featuring a separation larger by ~20 meV. **(c)** Schematic showing the optical transitions of spin-up excitons in $H_X^M$ hetero-bilayer at K valleys, where both $X_A^{Mo}$ and $X_B^{Mo}$ transitions are split into hybridized transitions ($X_{h+}^{\uparrow o,e}$ and $X_{h-}^{\uparrow o,e}$). Note that the valence band spin splitting in $H_X^M$ bilayer is increased from $2\lambda$ to $2\sqrt{\lambda^2 + t^2}$ by the presence of finite interlayer hopping ($t$), whereas in $R_M^X$ bilayer the interlayer hybridization is absent by symmetry. **(d,e)** Schematics showing the wavefunctions in the out-of-plane direction for the eight species of partially layer-hybridized excitons per valley with spin-up (d) and spin-down (e), in which four of them $X_h^o$ have larger optical dipoles (~97% compared to that of monolayer exciton) while four of them $X_h^e$ have larger electric dipoles (~97% compared to that of interlayer exciton).